\documentclass[utf8,cmex10]{frontiersSCNS} 

\setcitestyle{square} 
\usepackage{url,hyperref,lineno,microtype,subcaption}
\usepackage[onehalfspacing]{setspace}
\usepackage{amsmath}
\usepackage{amssymb}
\usepackage{threeparttable}
\usepackage{mathtools}

\DeclarePairedDelimiter\floor{\lfloor}{\rfloor}

\newcommand{\mean}[1]{\mathbb{E}[#1]}

\def\keyFont{\fontsize{8}{11}\helveticabold }
\def\firstAuthorLast{Inceoglu {et~al.}} 
\def\Authors{Fadil Inceoglu\,$^{1,2,3}$, N\'{e}stor J.\, Hern\'{a}ndez Marcano\,$^{1,*}$, Rune H.\, Jacobsen\,$^{1}$, and Christoffer Karoff\,$^{4}$}
\def\Address{$^{1}$Department of Engineering,  Aarhus University, Denmark \\
$^{2}$Cooperative Institute for Research in Environmental Sciences, University of Colorado Boulder, Boulder, CO, USA \\
$^{3}$National Centers for Environmental Information, National Oceanographic and Atmospheric Administration, Boulder, CO, USA \\
$^{4}$Department of Geoscience, Aarhus University, Denmark}
\def\corrAuthor{Department of Engineering,  Aarhus University, Finlandsgade 22, DK-8200 Aarhus N, Denmark}

\def\corrEmail{nh@au.eng.dk}

\begin{document}
\onecolumn
\firstpage{1}

\title[Localising SGRBs with a CubeSat mega-constellation]{A General Overview for Localizing Short Gamma-ray Bursts with a CubeSat Mega-Constellation} 

\author[\firstAuthorLast]{\Authors} 
\address{} 
\correspondance{} 
\extraAuth{}

\maketitle

\begin{abstract}

The Gamma-Ray Burst Monitor (GBM) on the {\it Fermi Gamma-Ray Space Telescope}, for the first time, detected a short gamma ray burst (SGRB) signal that accompanies a gravitational wave signal GW170817 in 2017. The detection and localization of the gravitational wave and gamma-ray source led all other space- and ground-based observatories to measure its kilonova and afterglow across the electromagnetic spectrum, which started a new era in astronomy, the so-called multi-messenger astronomy. Therefore, localizations of short gamma-ray bursts, as counterparts of verified gravitational waves, is of crucial importance since this will allow observatories to measure the kilonovae and afterglows associated with these explosions. Our results show that, an automated network of observatories, such as the Stellar Observations Network Group (SONG), can be coupled with an interconnected multi-hop array of CubeSats for transients (IMPACT) to localize SGRBs. IMPACT is a mega-constellation of $\sim$80 CubeSats, each of which is equipped with gamma-ray detectors with ultra-high temporal resolution to conduct full sky surveys in an energy range of 50-300 keV and downlink the required data promptly for high accuracy localization of the detected SGRB to a ground station. Additionally, we analyze propagation and transmission delays from receipt of a SGRB signal to ground station offload to consider the effects of constellation design, link, and network parameters such as satellites per plane, data rate, and coding gain from erasure correcting codes among others. IMPACT will provide near-real-time localization of SGRBs with a total delay of $\sim$5 s, and will enable SONG telescopes to join the efforts to pursue multi-messenger astronomy and help decipher the underlying physics of these events.

\section{}

\tiny
 \keyFont{ \section{Keywords:} Multilateration, short gamma-ray bursts, CubeSat, localization, constellation, communication, and TDoA}
\end{abstract}

\section{Introduction} \label{sec:intro}

The first detection of the gravitational wave (GW) signal from GW150914, which was produced by the mergers of stellar-mass black hole (BH) binaries, by the Laser Interferometer Gravitational Wave Observatory (LIGO) in 2015 \citep{2016PhRvL.116f1102A}, has opened a new era in GW astronomy. The LIGO consists of two observatories located in Hanford Site, WA, and in Livingston, LA, in the USA. The position of the detected GW signal source was calculated using time difference of arrival (TDoA) of the signals at the two detectors and it was localized to an area of 600 deg$^{2}$ in the sky \citep{2016ApJ...826L..13A,2016PhRvL.116x1102A}. Three years after its first detection, LIGO in the USA and Virgo, another observatory in Santo Stefano a Macerata, Italy, observed GW signals from GW170814. This helped the two teams to confine the position of the source to 60 deg$^{2}$ in the sky because the third observatory provided additional TDoA information \citep{2017PhRvL.119n1101A}.

In addition to the mergers of the stellar-mass BH binaries, the GWs can also be produced by the mergers of neutron star (NS) binaries and NS-BH binaries \citep{1991ApJ...380L..17P,2008PhRvL.100s1101A}, which are the most promising candidates for generating electromagnetic counterparts to the GWs \citep{2016ApJ...826L...6C}. These electromagnetic counterparts, which involve NSs, are proposed to be short gamma ray bursts (SGRBs).
The Advanced LIGO and Advanced Virgo gravitational-wave detectors have observed the first gravitational wave signal coming from a NS-NS binary spiralling in towards each other on 17th August 2017. Complementary to the detected GW signals, the Gamma-ray Burst Monitor (GBM) on {\it Fermi Gamma-Ray Space Telescope} detected a short gamma-ray burst, GRB 170817A, 1.7 s after the coalescence, supporting the first hypothesis of a neutron star merger. These subsequent detections made by gravitational wave and gamma ray observatories provided, for the first time, the direct evidence that merging neutron star binaries generate short gamma ray bursts as well as gravitational waves. Fast localization and identification of the electromagnetic counterparts enabled observations of the source across the whole energy spectrum from radio to gamma-ray wavelengths. This joint observational effort, so-called the multi-messenger astronomy, provides insights into astrophysics, dense matter, gravitation, and cosmology \citep{2017PhRvL.119p1101A}.

Gamma-ray bursts (GRB) are rapid and intense high energy prompt emissions peaking in the gamma-ray energies (hundreds of keV to MeV), with extended afterglows from radio to X-rays, GeV and even up to TeV gamma-rays \citep{2015JHEAp...7...73D, 2019ApJ...878...52A, 2019Natur.575..455M, 2019Natur.575..459M, 2019Natur.575..464A}. The durations of the detected GRBs show a bi-modal distribution, with local maxima at $\sim$0.2 s and $\sim$20 s and a transition around 2-3 s indicating two types of GRBs; (i) short gamma-ray bursts (SGRBs) with durations shorter than 2-3 s, and (ii) long gamma-ray bursts (LGRBs) with durations longer than 2-3 s \citep{2014ARA&A..52...43B,2015MNRAS.451..126S}. Together with their shorter durations, the SGRBs show almost no spectral lags and harder energy spectra compared with the LGRBs \citep{2015JHEAp...7...73D}. These observational differences suggest that the physical mechanism causing these SGRBs are of different origins than LGRBs. While the majority of nearby LGRBs are associated with core-collapse supernovae, mergers of the compact binaries, such as NS-NS or NS-BH, are proposed as the progenitors of the SGRBs \citep{2014ARA&A..52...43B,2015JHEAp...7...73D}.


Following the detection of a potential SGRB candidate, the communications payload of space-based gamma-ray observatories downlink the data for confirmation and analysis. The communications downlink conveys SGRB descriptive parameters as data packets transmitted through link and networking interfaces, whose end point are the mission control facilities on ground. Given the short time durations of SGRBs, the communications payloads of these missions are designed to ensure a low-latency downlink between the observatory and mission control to perform follow-up observations. Therefore, communications parameters such as link distance, packet size and data bit rates, which we refer simply as bit rate from now, are critical in determining the delay performance of a communications payload.

All space-based gamma-ray observatories, such as the \textit{Compton Gamma-Ray Observatory, the Fermi Gamma-Ray Space Telescope, and the Neil Gehrels Swift Observatory}, are very large and expensive satellite missions. As an emerging alternative platform, CubeSats can provide smaller, cheaper, and faster solutions \citep{2003ESASP.516..409T}. CubeSats arranged in a mega-constellation for global coverage, defined as a large number of CubeSats in various orbital planes that are synchronized and operate together, can increase the Field of View (FOV) to the whole sky and therefore can help increase the number of detected SGRBs. Such constellations introduce several communication hops as in the case of a mega-constellation used for detecting SGRBs. Further, they are also equipped with communication payloads fast enough to enable them to convey the potential GRB detection message within time frames similar to those of large and expensive observatory missions.

The message packets consist of sequential groups of bits known as symbols, which define the amplitude, frequency and phase of the wave signals transmitted. The received signals on ground used to reconstruct the data packets might suffer from absorption, refraction or dispersion at specific time, which affects the original bits inducing bit errors and making the data packets unusable. To counter this problem, the communications payload uses Forward Error Correction (FEC) \citep{hamming1950error}, which adds redundant bits that allow to detect and correct erroneous bits previously received. A packet loss, also known as a packet erasure, occurs when a packet arrives with internal errors at the bit level that cannot be corrected even after employing error correcting codes. In this case, the whole packet is discarded, even if some sections of it are useful. A packet loss can also be considered as having occurred when packets do not arrive at all when supposed to. For all these cases, a new type of FEC is implemented above the bits at the packet level known as erasure correcting code. This is achieved by creating redundant packets (instead of redundant bits) as combinations of original ones, which translates in more transmissions utilized to recover when packet losses occur. Thus, erasure correcting codes ensures all data packets are correctly received. However, there is a trade-off between the tolerable delay and the amount of redundant transmissions needed for protection. This depends on the needs of the given mission scenario and is therefore of great interest to search for codes that minimize their introduced redundancy.

Among the alternatives for erasure correcting codes as FEC, block codes such as Reed-Solomon \citep{ReedSolomon1960polynomial}, Low Density Parity Check (LDPC) \citep{1963gallagerldpc}, or rateless codes such as Raptor \citep{shokrollahi2006raptor} could be utilized to cope with packet losses. However, these codes have the drawback of working on a point-to-point basis. This implies that a set of packets needs to be encoded and decoded for every time they are sent between a transmitter and a receiver. Codes based on point-to-point communication incur in delays which are critical for transient applications, given the need of several hops before reaching the ground system for analysis. For this case, Network Coding (NC) \citep{ahlswede2000network} and particularly RLNC \citep{ho2006random} provides an advantage against traditional codes since it does not require to encode and decode on a hop basis. Instead, RLNC sends coded packets as linear combinations of the original set, removing the need to get each individual packet by conveying linearly independent coded packets. This has also the advantage of not acknowledging each single packet, but once the whole set is decodable \citep{sundararajan2008arq}. Thus, RLNC coded packets can be forwarded faster in a potential multi-hop network to reduce the transmission delay when considering packet losses.

Here we will calculate, using a first order approach, the average number of CubeSats required in an interconnected multi-hop array of CubeSats for transients (IMPACT) to localize a SGRB with high-accuracy and the time it takes for the signal to be transmitted to a ground station on Earth. We assume that each CubeSat in the mega-constellation carries a gamma-ray detector for energies ranges of 50-300 keV, with an effective area between 400 - 600 cm$^{2}$ mounted on the largest side surface (20 cm $\times$ 30 cm) of the CubeSat. Nanosatellites will orbit the Earth in the Low Earth Orbit (LEO), which present the advantage of reduced propagation delays when compared to their Medium Earth Orbit (MEO) and Geostationary Orbit (GEO) counterparts.

The SGRBs are most likely associated with GWs, and an accurate and prompt localization of the source of these events will enable us to perform follow-up measurements of the afterglows using an automated network of ground-based observatories such as SONG. As an automated network of observatories, SONG can perform photometric and spectroscopic measurements. The FOV of the SONG telescope is 30$''$\,$\times$\,20$''$ and it is possible to control the telescopes remotely.

In this work, we define a mega-constellation of CubeSats to detect SGRBs, present a discussion about ideal payloads, and analyse the communication parameters for downlinking the data promptly. Our contributions can be listed as follows:

\begin{itemize}
\item We review the properties of state-of-the-art scintillation crystals and their readout electronics to ensure we meet the energy range, light yield, decay time, and time resolutions to localise SGRBs with high-accuracies. 
\item We calculate the number of CubeSats required to localise SGRBs with our proposed guidelines, regardless of their origin in space, and for a given accuracy and time uncertainty based on the geometrical properties of the detecting constellation.
\item We perform a timing analysis in terms of wave propagation time and data transmission time to observe the effects of the communication system parameters in the event report delay, where we propose RLNC as a suitable erasure correcting code for this mega-constellation~\citep{hernandez2019delay}. Our analysis reveals that the interplay between constellation and communication parameters allows to reduce the total communication delay, reaching performance metrics with similar order of magnitude of scientific space missions. We also find that ideal constellation configurations that minimise the total communication delay for the worst case scenario existing for a given set of parameters. To the best of our knowledge, this is the first time that such analysis is presented.
\end{itemize}

Our work is organised as follows: Section~\ref{sec:Cube} describes the CubeSat technology, trends, and prior work regarding detection of SGRBs. In Section~\ref{sec:GammaDet}, we present our guidelines for gamma ray detectors, where the localization method is described in Section~\ref{sec:Trian}. The communications scheme that each CubeSat has to convey the event to ground is presented in Section~\ref{sec:DatTrans}. Section~\ref{sec:ConstDes} indicates how the constellation is constructed to ensure that we reach a given total number of satellites for the detection requirements. In Section~\ref{sec:AnRes}, we present our analyses and results for the required number of satellites for a given localization accuracy and time delay uncertainty, and the total delay for a given set of communication parameters and different configurations. Section~\ref{sec:conc} presents the discussion, and conclusions of our work.

\section{CubeSats}
\label{sec:Cube}

A CubeSat, which was developed as a collaborative effort between California Polytechnic State University and Stanford University's Space Systems Development Laboratory in 1999 \citep{2003ESASP.516..409T}, is a standardized model of a miniaturized satellite with a volume of $10\times10\times10$ cm$^{3}$ and a maximum weight of 1.33 kg (1U CubeSat).

CubeSat Data\footnote{Saint Louis University CubeSat Database: {\url{https://sites.google.com/a/slu.edu/swartwout/home/cubesat-database\#refs}}} since the mid-2000s shows that the number of CubeSat missions based on one or several 1U architectures have been increasing, where it showed a sudden jump in 2013. Starting from 2014, most of the CubeSat missions have been mainly on Earth observations and technology demonstrations, while technology demonstration missions have been decreasing since 2013. Scientific missions on the other hand gained momentum in 2017. Major factors in the increase of these type of missions have been the reduction in launch costs and the reliability of spacecraft of subsystem components to space applications\footnote{Erik Kulu. Nanosats EU: \url{http://nanosats.eu/}}\footnote{Erik Kulu. Newspace Index: \url{https://www.newspace.im/}}.

There are several CubeSat missions planned for detection and localization of short- and long-GRBs, former of which can be associated with GW signals, using 5U \citep{Yonetoku2014} and 6U \citep{2017arXiv170809292R} CubeSat architectures, as well as a swarm of 3U CubeSats \citep{2018SPIE10699E..64O}. To localize the SGRBs accompanying the GWs, \citet{Yonetoku2014} developed Kanazawa-SAT, an X-ray imager based on a coded masked-silicon drift detector, size of which is 100 cm$^{2}$, mounted on a 5U CubeSat. The FOV of the developed X-ray detector is $\sim$1 steradian and therefore \citet{Yonetoku2014} suggested to use a fleet of various Kanazawa-SATs to reach whole sky (4$\pi$ steradians) coverage. On the other hand, \citet{2017arXiv170809292R} presents BurstCube, a design of a 6U CubeSat carrying 4 CsI scintillators coupled with arrays of Silicon photomultipliers (SiPMs) to complement the bigger missions like {\it Fermi Gamma-Ray Space Telescope} and Neil Gehrels {\it Swift} Observatory, but also capable of detecting and localizing the SGRBs. The final proposed configuration of BurstCube includes a constellation of $\sim$10 6U CubeSats to provide whole sky coverage. \citet{2018SPIE10699E..64O} designed CubeSats Applied for MEasuring and LOcalising Transients (CAMELOT), a fleet of 9 3U CubeSats each carrying a CsI(TI) scintillator coupled with two Multi-Pixel Photon Counters (MPPC) as read-out electronics to detect a localize SGRBs. Based on this architecture, \citet{2018SPIE10699E..64O} calculated localization accuracies better than 20$'$.
However, all these prior missions or designs only allow to provide at most coverage when a GRB event occurs in the same hemisphere as the CubeSats' current transit. If this is not the case, then the opportunity for detection and analysis will be lost. Further, even if a GRB event is detected in a given hemisphere the same as the transit of various CubeSats, but no nearby ground stations exist, then there will be no connectivity possible to provide a low-latency response from detection and reporting.

\section{Gamma-ray detectors}
\label{sec:GammaDet}

The most commonly used gamma-ray detectors are made of scintillator detector materials, such as Thallium-doped Sodium Iodide (NaI(Tl)), Thallium-doped Cesium Iodide (CsI(Tl)), and a relatively new Cerium-activated Lanthanum(III) Bromide (LaBr$_{3}$(Ce)) which provides a higher energy resolution than the previous two \citep{2008pgrs.book.....G}. Prices for a type of a single scintillator\footnote{Saint Gobain Crystals: {\url{https://www.crystals.saint-gobain.com/products/standard-and-enhanced-lanthanum-bromide}}} are on the order of $\sim$EUR 40,000. Such detector, including readout electronics, is able to fit in a 3U section of a 6U CubeSat leaving space for other essential subsystems (power supply, communications, on-board computer, etc). In a scintillation crystal, incoming gamma-rays produce primary electrons, which lose their energy by creating secondary electron-hole pairs in the crystal lattice. The created secondary electrons can be found in excited state, which then de-excite by emitting electromagnetic radiation. If this radiation is in, or near, optical wavelengths, it will be detected by a photomultiplier that generates an electrical signal to provide the detector signal. The decay time of the scintillation crystal, which is defined as time it takes for an excited secondary electron to de-excite, must be short to allow high count rates \citep{2008pgrs.book.....G}. Higher counting rates will, in turn, help reduce the uncertainty in detection of timing of a SGRB.

\begin{table*}[ht!]
\centering
\begin{threeparttable}
\caption{Properties of most common scintillator materials for gamma-ray detection.} 
\label{tab:Sint}
\begin{tabular}{lccccc}
\hline
\hline
\multicolumn{1}{l}{Scintillator} 	& \multicolumn{1}{c}{light yield} 	&  \multicolumn{1}{c}{decay}	& \multicolumn{1}{c}{density} 		& \multicolumn{1}{c}{refractive} & \multicolumn{1}{c}{wavelength at}  \\ 
\multicolumn{1}{l}{} 			& \multicolumn{1}{c}{(photon/keV)}  	& \multicolumn{1}{c}{time (ns)} 	& \multicolumn{1}{c}{$\text{g\,cm}^{-3}$}  	& \multicolumn{1}{c}{index} 	 & \multicolumn{1}{c}{max. emission (nm)} \\
\hline

LaBr$_{3}$(Ce)$^{a}$ 			& 63  						& 16							& 5.08						& $\sim$1.9 					& 380\\
CsI(Tl)$^{b}$ 					& 52  						& 1000						& 4.51	 					& 1.79						& 550\\
NaI(Tl)$^{b}$ 					& 38  						& 230						& 3.67 						& 1.85						& 415\\
\hline
\end{tabular}
\begin{tablenotes}
\item [a] Data in the related rows are taken from LaBr$_{3}$(Ce) data sheet in https://www.crystals.saint-gobain.com
\item [b] Data in the related rows taken from \citep{2008pgrs.book.....G}.
\end{tablenotes}
\end{threeparttable}
\end{table*}

Among the three most common scintillator materials, LaBr$_{3}$(Ce) provides the shortest decay time of 16 ns (Table~\ref{tab:Sint}). The decay time depends on the concentration of cerium as an activator in the LaBr$_{3}$ crystal. This value is crucial as reaching high localization accuracies is correlated with high photon counting rates for a given time interval (bin). Having ultra-high temporal resolution of gamma-ray detection will increase the photon counting rate, and hence provide better counting statistics. LaBr$_{3}$(Ce) has 40\% and 17\% higher light yield, which is defined as the efficiency in converting ionization energy to light output in the scintillation crystal, compared with NaI(Tl) and CsI(Tl), respectively. LaBr$_{3}$(Ce) has an inherent radioactive impurity caused by a 0.09\% unstable $^{138}$La isotope from the default lanthanum, and contamination from its homologue $^{227}$Ac \citep{2007NIMPA.574..115Q,2007NIMPA.582..554N}. However, crystal processing refinements reduced this contamination by a factor of 15 \citep{2007NIMPA.574..115Q}. This intrinsic activity of the LaBr$_{3}$(Ce) crystal is not expected to impact the ultra-high temporal resolution, although it might increase the background counts \citep{2007NIMPA.574..115Q}. LaBr$_{3}$(Ce) also has a higher efficiency in stopping power due to its higher density (Table~\ref{tab:Sint}), meaning higher counts in full energy peaks.

In addition to light yield, decay time and density, refractive index and wavelength at maximum emission are also important features of a scintillator material for choosing the most compatible read-out electronics, such as photo-multipliers. Photo-multipliers convert the output of the scintillation crystal, which is a quantity of light, into an electrical signal \citep{2008pgrs.book.....G}. Conventionally, this is achieved by using photomultiplier tubes (PMTs). PMTs require high biasing voltages as well as they are bulky and fragile and sensitive to magnetic fields \citep{2018ITNS...65..645C}. Silicon photomultipliers (SiPMs), on the other hand, are relatively new technology. SiPMs are made of dense arrays of avalanche photodiodes (APDs), and they are insensitive to magnetic fields, they have high multiplication gain ($\sim$10$^{6}$) and hence negligible electronic noise, low operation voltages, compact designs, and they provide very high timing resolution in the order of a few nanoseconds \citep{2015JPhCS.620a2001J,2018ITNS...65..645C,Butt2015}. In addition to the SiPMs, Silicon Drift Detectors (SDD) as read-out apparatus of scintillators provide high quantum efficiency ($>$80\%) and no photo-multiplication process. However, having no photo-multiplication process makes the system vulnerable to the readout electronic noise \citep{Butt2015}. Although SDDs provide much better spectral resolutions, their drift time, which is defined as the time delay between the time the charge carriers are produced inside the detector and the time that the corresponding current peak at the output is observed, poses a problem as it is generally in the order of a few microseconds \citep{Butt2015}. 

Depending on the main purpose of the gamma-ray detector, the two photo-detectors types could provide different solutions; (i) SiPMs for a higher time-resolution, and (ii) SDDs for higher spectral resolution (see Table 1 in \citep{Butt2015}). Noise contribution due to the dark current in the SiPMs, which is defined as the excess leakage current of a photodiode in reverse bias in the absence of light, is a known issue. However, this issue might be overcome by using SiPMs with small surface areas as dark current scales linearly with their surface area \citep{2018JInst..13P6001B}, and a coincidence readout method \citep{2018SPIE10699E..64O}. Another known issue is the radiation damage in SiPMs, however the effects of low energy photons ($\leq$ 300 keV) are limited to surface damage instead of the bulk material caused by higher energy photons \citep{mitchell2020radiation}. Therefore, using SiPMs photo-detectors as read-out electronics coupled with a LaBr${}_3$(Ce) scintillator crystal will enable us to obtain a fast response and high temporal resolution, which will in turn help increase the photon counting rates. 

In this study, we consider that the design, fit, and volume of our gamma-ray detectors along with their passive shielding to reduce the effects of scattered gamma-rays, are planned to be in line with the previously proposed missions, such as the BurstCube \citep{2017arXiv170809292R} and CAMELOT \citep{2018SPIE10699E..64O}. This will guide us to avoid exceeding the volume and weight limits, and reduce the effects of gamma ray scattering for our gamma-ray detectors that will be mounted on our planned 6U CubeSats. The effective area of our gamma-ray detectors for the energy range 50 - 300 keV is planned to be between 400 to 600 cm$^{2}$, which could be mounted on the largest side surface (20 cm $\times$ 30 cm) of a 6U CubeSat. The large effective area is expected to increase the photon counting rates, and hence improve the counting statistics.

\section{Triangulation of SGRBs}
\label{sec:Trian}

\begin{figure}[htb!]
\begin{center}
{\includegraphics[width=2.5in]{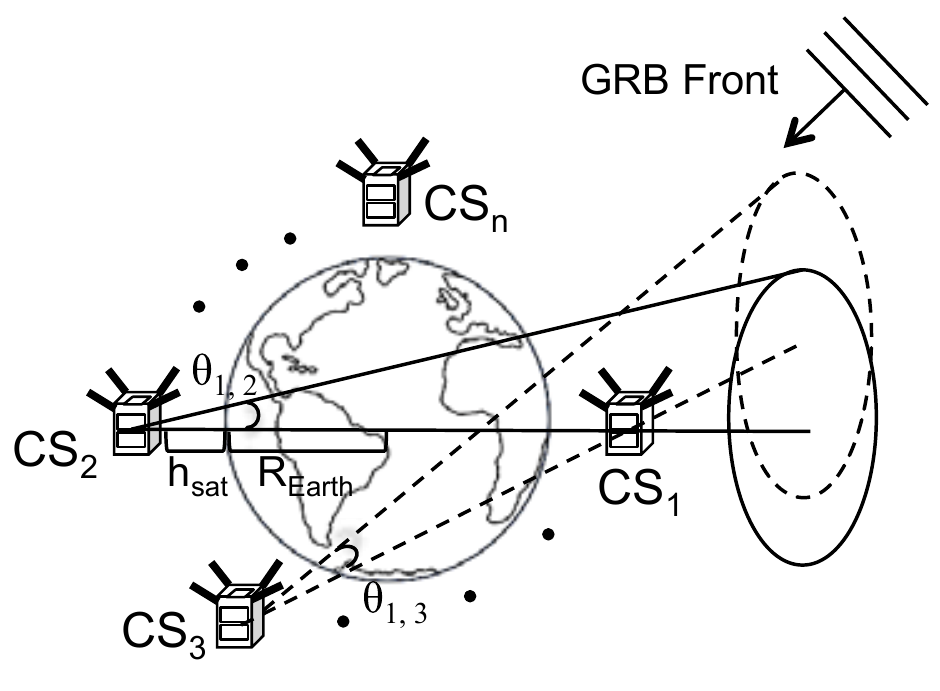}}
\caption{Illustration of the triangulation method using CubeSats. Each independent CubeSat pair is used to derive an annulus of location for the GRB source.} 
\label{fig:Trian}
\end{center}
\end{figure}

Localization of the GRB sources in space is planned to be achieved by triangulation method (Figure~\ref{fig:Trian}). In this method, when a bright GRB occurs in deep space, the photons coming from the source are detected by the first CubeSat (CS$_{1}$) at time $t_{1}$, while they are registered by the second CubeSat (CS$_{2}$) at the time $t_{2}$.

Assuming that the GRB is a planar wave, meaning that the distance to the source of the event is much larger than the distance between the two CubeSats, the direction of the GRB source can be constrained by the triangulation formula \citep{2012JPhCS.363a2034P,2013ApJS..207...39H},

\begin{eqnarray}
\cos \theta_{1,2}=\frac{c \, (t_{2}-t_{1})}{D_{1,2}}=\frac{c \, \delta t}{D_{1,2}}
\label{eq:TriLat1}
\end{eqnarray}

\noindent where $\theta_{1,2}$ is the half-angle of the annulus with respect to the vector joining CubeSats 1 and 2, $c$ is the speed of light, $D_{1,2}$ is the distance between the two CubeSats, and $\delta t$ is the time delay of arrival of the photons between the two CubeSats. A third CubeSat jointly with the previous two produces two possible error boxes causing ambiguity in localization. The ambiguity problem can be overcomed by using a fourth CubeSat (or more) in a non-coplanar orbit \citep{2013ApJS..207...39H}. The finite width of this annulus $d\theta_{1,2}$ and one dimension of the resulting error box $\sigma(\theta_{1,2})$, can be calculated by propagating the uncertainty (error) of the time delay in Equation~\ref{eq:TriLat1} as follows \citep{2012JPhCS.363a2034P,2013ApJS..207...39H},

\begin{eqnarray}
d\theta_{1,2}= \sigma(\theta_{1,2}) = \frac{c \, \sigma(\delta t)}{D_{1,2}\, \sin \theta_{1,2}}
\label{eq:TriLat2}
\end{eqnarray}

\noindent where $\sigma(\delta t)$ is the uncertainty in the time delay. The radius of each annulus and the right ascension and declination of its centre are calculated in a heliocentric frame. The time delay of arrival of the SGRB signal from two satellites $\delta t$, can be calculated using the cross correlation method of the observed light curves, which requires precise time synchronizations between the satellites. The uncertainty in the time delay calculations is directly linked to the binning time for count reporting, the photon counting rates and hence, a gamma ray detector with a large effective area must be considered to decrease the uncertainties in counting statistics \citep{2018SPIE10699E..64O}. Here we consider an effective area between 400 to 600 cm$^{2}$ for energies ranging from 50 to 300 keV, corresponding to the largest side surface (20 cm $\times$ 30 cm) of a 6U CubeSat architecture.

Historically, triangulation for localizations of GRBs has been performed by the Interplanetary Network (IPN) \footnote{\url{https://heasarc.gsfc.nasa.gov/docs/heasarc/missions/ipn.html}}. The $3^{rd}$ IPN is in operation since the launch of \textit{Ulysses} in 1990 \footnote{\url{http://www.ssl.berkeley.edu/ipn3/index.html}}, and currently consists of \textit{Konus-Wind} in a heliocentric orbit at L1 point between Earth and Sun \citep{1995SSRv...71..265A}, \textit{Mars Odyssey} orbiting Mars \citep{2006ApJS..164..124H}, \textit{the International Gamma-Ray Laboratory (INTEGRAL)} in GEO \citep{2005A&A...438.1175R}, \textit{Neil Gehrels Swift Observatory} \citep{Gehrels_2004} in LEO, \textit{Fermi} \citep{Meegan_2009} in LEO, and \textit{BepiColombo}, which will arrive in an orbit around Mercury in late 2025 \footnote{\url{https://sci.esa.int/web/bepicolombo}}. The farthest member of the $3^{rd}$ IPN is Mars Odyssey that provide maximum baseline distance of $\sim$2.52 AU, when Earth and Mars are the farthest apart. This advantage however comes with a cost of prolonged signal transmission times of about 21 minutes.
      
\section{System Overview \& Data Transmission Scheme}
\label{sec:DatTrans}

\begin{figure}[htb!]
\begin{center}
{\includegraphics[width=7in]{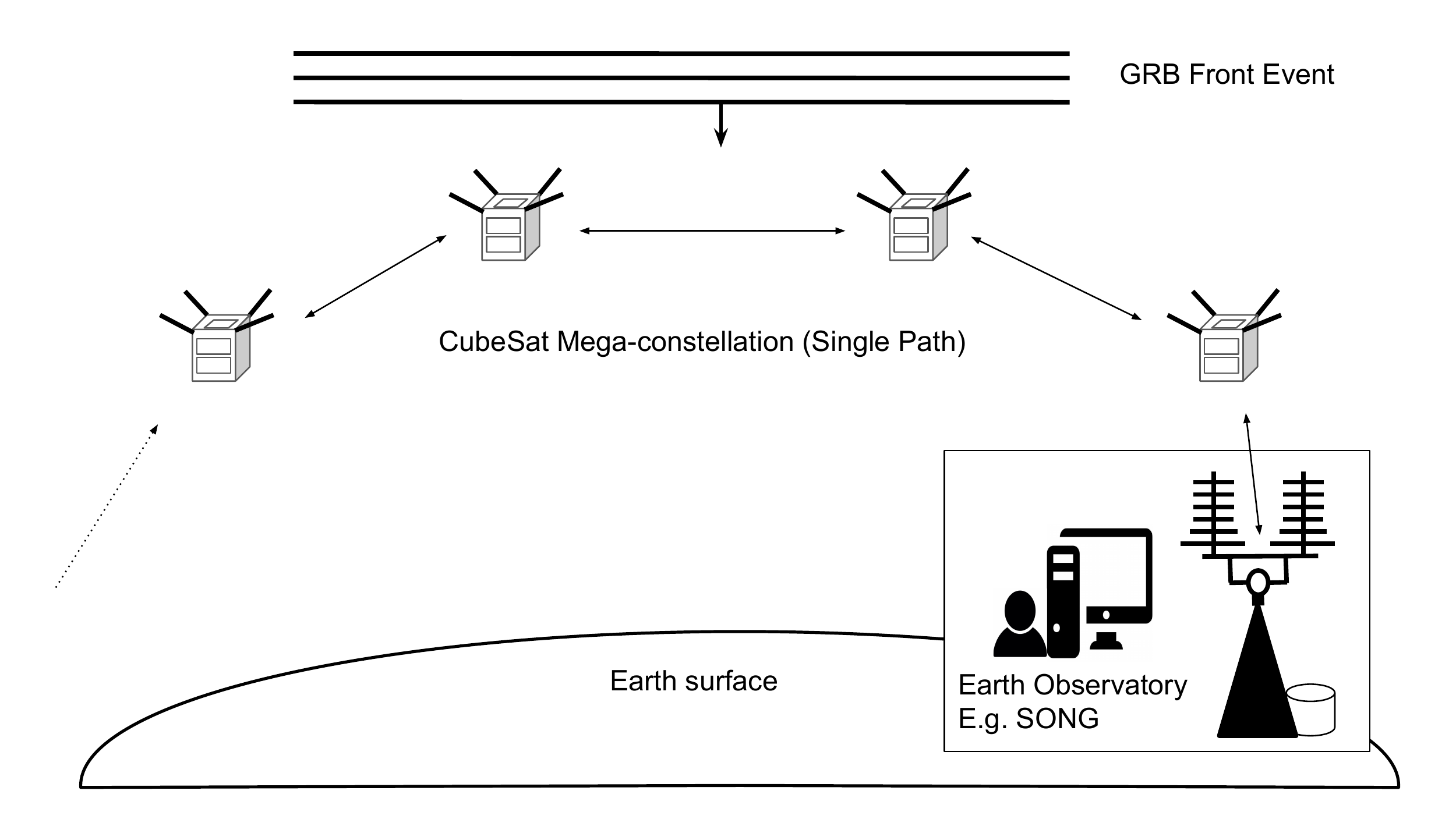}}
\caption{IMPACT system overview of one possible data transmission path in a CubeSat mega-constellation detecting a GRB front from above. The path consists of a various detecting and data forwarding CubeSats to an Earth observatory for analysis, e.g. a ground station in SONG.} 
\label{fig:impactsyst}
\end{center}
\end{figure}

Performing a fast triangulation of GRBs with mega-constellation of CubeSats in LEO requires to define a distributed multi-hop network, given the limited access time to a CubeSat from ground. Such a constellation allows to establish various paths from the detecting CubeSats to the different ground stations of SONG. In Figure~\ref{fig:impactsyst}, we present a system overview of an example path in such network. The path consists of four CubeSats for detection and providing connectivity to a ground station of the SONG ground network. After the GRB front has been detected by a CubeSat, a transmission with descriptive data of the detected light curve event is sent across the path. Then, each CubeSat forwards the data across the network until reaching the ground station belonging to the SONG network. To reach this goal, we consider our transmission scheme in a given established physical network with: (i) available energy subsystem budget, (ii) radio communications payload in the satellite, (iii) access scheme for communication resource (frequency and timeslot) allocation, (iv) corresponding network addressing space, and (v) static routing, which we all consider assume fixed and operative. This ensures each data packet (referred as segment at transport) to be properly routed by a CubeSat, and also acknowledged both after transmitted and once successfully received by the ground station. Given that we design our constellations for global coverage, there will always be a connection with the ground station and a path will be established to it. At the end, all the light curve events are gathered in the same facility (e.g. a control room) for data analysis, estimation, and visualization of the relative position and time of the GRB event through the triangulation method described earlier. 

Mega-constellations benefit from shorter round trip times (RTT) between two neighboring satellites in the network. We define the RTT as the sum of times taken for: (i) the propagation of the signal from a transmitting node in the network to a receiving node in a single hop, and (ii) the propagation of the corresponding acknowledgement signal in the opposite direction. Thus, the RTT for a 580 km LEO orbit that we consider, is $\sim$3.87 ms. This is smaller compared to those in the geostationary orbits, which have RTTs of approximately 240 ms. The shorter RTT of LEO compared to Medium Earth Orbit (MEO) and Geostationary Orbit (GEO) satellite systems, makes LEO satellite networks better in time efficiency when signal transmissions among satellites are considered. Achieving fast and reliable transmission of mission-related data will likely enable near-real-time detection and localization of the SGRBs.

In our proposed system constellation, we consider a reliable data transmission scheme for each hop from a detecting CubeSat to a ground station collocated at SONG. Data transport is based on an end-to-end protocol scheme from a networking perspective, while we consider this in each single possible detection path. However, for our timing analysis, we consider the longest possible time delay occurring when a detection data packet transmission goes through the longest path in the constellation, since all other paths will incur in smaller delays, and thus, will be upper bounded by the delay of the longest path over several hops.

Multi-hop networks, however, might experience segment loss, which in turn affects the performance of the end-to-end protocol scheme \citep{sundararajan2011network}, increasing the total delay which is critical in our case. A hop-by-hop version of the Reliable Datagram Protocol (RDP) \citep{rfc908} is utilized in these type of networks and could reduce this loss. However, RDP establishes a handshake at each hop-pair which will add further time-delays. To minimise the data segment loss rates and the RTT at each hop in the mega-constellation, we use Random Linear Network Coding (RLNC) \citep{ho2006random} at the transport layer, which allows the data segments to be reliably transmitted despite segment losses \citep{hernandez2019delay}. This represents the case of FEC in terms of an erasure correcting code at the transport layer. Using RLNC avoids encoding and decoding at each immediate hop, in contrast to the other rate-based or rate-less block codes at the transport layer \citep{kim2012coded}. With RLNC, segments are acknowledged as a set on a hop basis removing the caveats of RDP.  In this way, RLNC can be regarded as a reliable version of the User Datagram Protocol (UDP) \citep{rfc768}, but without a flow or congestion control mechanism. All these benefits allow RLNC to achieve a lower latency when compared with RDP, which is critical to fetch event reports for analysis.

RLNC achieves these benefits by splitting the transmitted data into blocks of equal size called generations, where each consists of $g$ segments. In each block, coded segments are created as linear combinations of the segments within that block, where the multiplying coding coefficients are drawn from a Galois Field (GF) of size $q$. In RLNC, it is possible to generate recoded segments by recoding previously received coded segments. This process is called coding online or on-the-fly, where decoding coded segments prior to forwarding them to the next hop is not necessary \citep{sundararajan2008arq}. To decode a generation, only a linearly independent set of coded segments is necessary and sufficient to recover the original segments. Any coded segment gets appended to its header a total amount of bits equal to $\alpha = g\log_2(q)$ as overhead to indicate which coding coefficients were used to create them \citep{heide2011code}, which are necessary in the decoding process. In our system, the decoding is performed through Gaussian elimination at the final receiver, the ground station, as it possesses the required computation power.

\begin{figure}[htb!]
\begin{center}
{\includegraphics[width=3in]{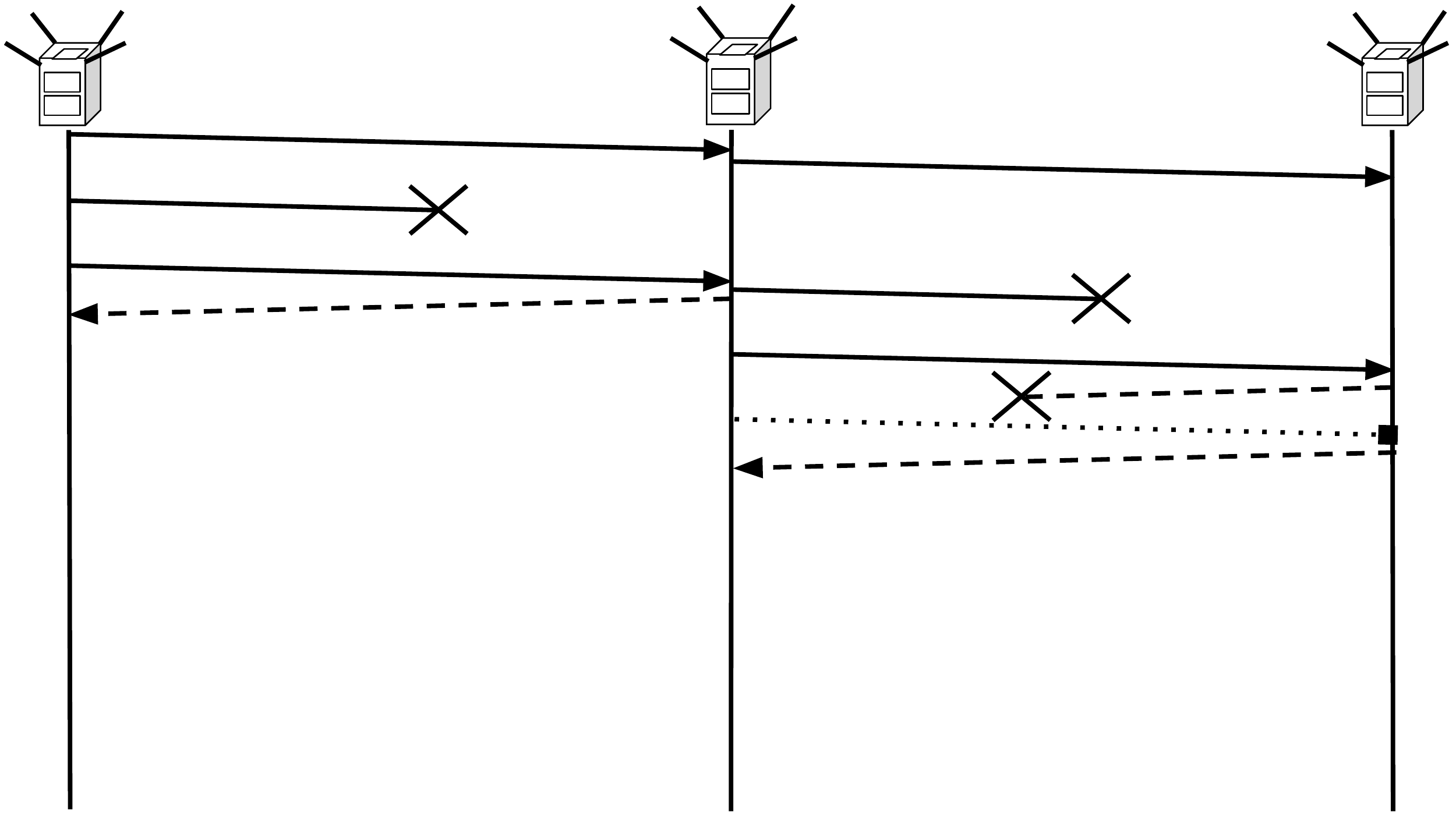}}
\caption{Illustration of the RLNC for the case of two segments and a two hops. The vertical lines represent increasing time from top to bottom and we assume a high field size in for the sake of simplicity. The solid arrows represent coded segments, whereas the dashed arrows represent acknowledgements.} 
\label{fig:TRLNC}
\end{center}
\end{figure}

An example of coping with segment losses with RLNC  as FEC is shown in Figure~\ref{fig:TRLNC}. The vertical lines represent increasing time from top to bottom. At the beginning of the transmission, the leftmost CubeSat starts transmitting coded segments that can be received (solid arrow in Figure~\ref{fig:TRLNC}) or lost (solid with cross in Figure~\ref{fig:TRLNC}). Segments are sent continuously until enough linearly independent segments are received and an acknowledgement is transmitted back for the sending to stop (dashed line). The middle CubeSat starts transmitting towards the CubeSat on the right as soon as segments are received from the previous hop if they are linearly independent. This process is repeated across all hops until the whole message is received at a ground station to decode the data. This is in contrast with any other type of necessary FEC, where all segments would have to be decoded in each hop before being coded and sent to the next introducing more delays.

\section{Constellation Design}
\label{sec:ConstDes}

To detect and localise SGRBs with high accuracies, we consider using a total number of $N_{sat}$ CubeSats in the LEO for continuous Earth coverage, which are time-synchronised and operate together. An interconnected distributed network of CubeSats in a mega-constellation is therefore composed of $P$ number of planes (polar orbits), each of which accommodates $S$ CubeSats making $P \times S = N_{sat}$, and where various configurations are possible for full sky coverage. This constellation design is classified as Walker type \citep{1971JBIS...24..369W}. Such polar orbits at 580 km (as the \textit{Fermi Gamma-Ray Space Telescope}) should not affect the sensitivity performance with respect to background, scintillator activation and SiPM radiation damage, since collision effects from e.g., the inner Van Allen radiation belt, do not appear yet at this altitude and inclination. As an example, the Complex Orbital Observations Near-Earth of Activity of the Sun - F  (CORONAS-F) satellite \citep{KUZNETSOV20021857} operated properly in its complete lifetime under similar energy ranges, orbit height and inclination. The constellation is also designed to ensure that satellites do not collide at the poles. This is reached by allowing relatively small differences in the orbital inclinations and heights without affecting our analysis. From an operational and maintenance perspective, satellites will be deployed similarly to commercial missions, i.e. progressively set in orbit in batches, in their planned natural orbits until the final arrangement for start of operations is reached. The satellites in our work do not consider station keeping, nor propulsion systems, and therefore translation and attitude are determined by natural orbital mechanics. The constituent satellites of the constellation will decay naturally by Earth's gravitational pull and disintegrate on Earth's atmosphere progressively, where replacements will be approached on a failure basis as needed. The total number of satellites depends on a target localization accuracy and available delay uncertainty, which we review in Section~\ref{sec:AnRes}. For a given total number of satellites $N_{sat}$, we review the different possible configurations of planes and satellites per plane. Based on this, the intra-plane distance, defined as the distance between the CubeSats in the same plane, is given by \citep{ekici2001adistributed},
\begin{eqnarray}
d_{intra} = (r_E + h) [2\,(1 - \cos\left(2\pi/S\right))]^{\frac{1}{2}}
\label{eq:intrasat_dist}
\end{eqnarray}
\noindent where $r_E$ is the Earth radius and $h$ is the orbit height for all satellites. The orbit height is selected to reach a balance between total number of satellites and delay. Additionally, the inter-plane distance, which is defined as the distance between CubeSats in adjacent planes, can be calculated following \citep{ekici2001adistributed},
\begin{eqnarray}
d_{inter} = (r_E + h) [2\,(1 - \cos\left(\pi/P\right))]^{\frac{1}{2}}\times \cos \phi
\label{eq:intersat_dist}
\end{eqnarray}
\noindent where $\phi$ is the latitude at which a series of aligned satellites from different planes are located. The distance between a satellite and the ground station, to which the data will be transmitted is given as \citep{ekici2001adistributed}

\begin{eqnarray}
d_{gnd} = [r_E^2 \sin^2 \varepsilon_{min} + 2 r_E h + h^2]^{\frac{1}{2}} -r_E \sin \varepsilon_{min}
\label{eq:gnd_dist}
\end{eqnarray}

\noindent where $\varepsilon_{min}$ is the minimum elevation angle. The effect of the elevation angle will be discussed in Section~\ref{sec:AnRes}. As the orbital planes are assumed to be circular, these distances will be constant throughout the mission lifetime permitting to define the delay between any given pair of satellites across various hops.  

\section{Analyses and Results}
\label{sec:AnRes}
\subsection{Number of CubeSats and accuracy}

In this study, we aim to calculate a number of 6U CubeSats to achieve a high-accuracy localization so that SGRBs after-glows can be observed by SONG. The desired localization accuracy for the SGRBs must be much smaller than 20$''$. To achieve this objective, we assume that:

\begin{itemize}
\item Each CubeSat in a mega-constellation is equipped with gamma-ray detectors for the proposed	energy range between 50-300 keV, in compliance with the energy range of the nominal BATSE on-board burst trigger \citep{1989dots.work...96F},
\item The accuracy of the detection timing of a SGRB is correlated with the total number of photons counted during its duration. Considering that the median photon fluxes in 50 - 300 keV energy range for the SGRBs are $\sim$2 photon\,cm$^{-2}$\,s$^{-1}$ \citep{2020ApJ...893...46V}, the effective surface area of the gamma-ray detectors for this energy range must be between 400 to 600 cm$^{2}$, corresponding to the side surface of a 6U CubeSat architecture (20 cm $\times$ 30 cm). A larger effective surface area will increase the accuracy in detecting the timing of the GRB trigger,
\item When an increase in photon counting rates per time bin exceeds 5$\sigma$ threshold above background, it will be accepted as a SGRB trigger,
\item The average number of triggers per year is estimated based on the integral distribution of SGRB fluence in the energy range 50 - 300 keV \citep[the bottom panel of Figure 10 in][]{2020ApJ...893...46V}. We calculate the average energy fluence by

\begin{eqnarray}
\overline\Psi  = \overline E \times \int_{t_{0}}^{t_{1}} \phi (t) dt
\label{eq:Gen}
\end{eqnarray}

\noindent where $\overline E$ is geometric average of the energies 50 and 300 keV and the integral term is the total photon counts for the average duration of a SGRB. For SGRBs with photon fluxes larger than 5 photon\,cm$^{-2}$\,s$^{-1}$, we estimate the average number of SGRB detection as around 20 year$^{-1}$. As for photon fluxes more than 10 photon\,cm$^{-2}$\,s$^{-1}$, this rate is around 9 SGRBs year$^{-1}$.

\item The gamma-ray detectors will have time resolutions in the order of $\mu$s, which is linked to the read out electronics, to time stamp every photon with its precise arrival time. Higher resolution in time will lead to higher counting rates when time-binning the data leading to reduction of uncertainties in calculated time lags based on cross correlation,
\item We take into account only the statistical error related to $\delta$t and neglect uncertainties in distances between each pair of CubeSats, as the main contribution comes from the timing uncertainties \citep{2012JPhCS.363a2034P}. 
\item All of the CubeSats in the mega-constellation must be time-synchronized and their position must be known precisely, which can be achieved with the GPS technology. It was shown that synchronizing a GPS with a CsI scintillation crystal gamma-ray detector and SiPM as photo-detector is possible and this setup can provide GPS time stamping of the incoming gamma-ray photons with accuracy and precision better than $\sim$20 $\mu$s \citep{2018arXiv180603685P}.
\end{itemize}

The average localization accuracy for the first order approximation can therefore be written as:

\begin{eqnarray}
d\theta \approx \frac{c}{(r_E+h)} \frac{\overline{\sigma}(\delta t)}{\sqrt{N_{sat}-1}}
\label{eq:Gen}
\end{eqnarray}

\noindent where $\overline{\sigma}(\delta t)$ is the uncertainty in the average time delay among the paired-CubeSats, ($r_E+ h$) represents the average baseline, where $r_E$ is the Earth's radius, $h$ is the orbital altitude, $c$ is the speed of light and $\sqrt{N_{sat}-1}$ is the factor that controls the statistical improvement in determining the localization accuracy.

\begin{figure}[htb!]
\begin{center}
{\includegraphics[width=6in]{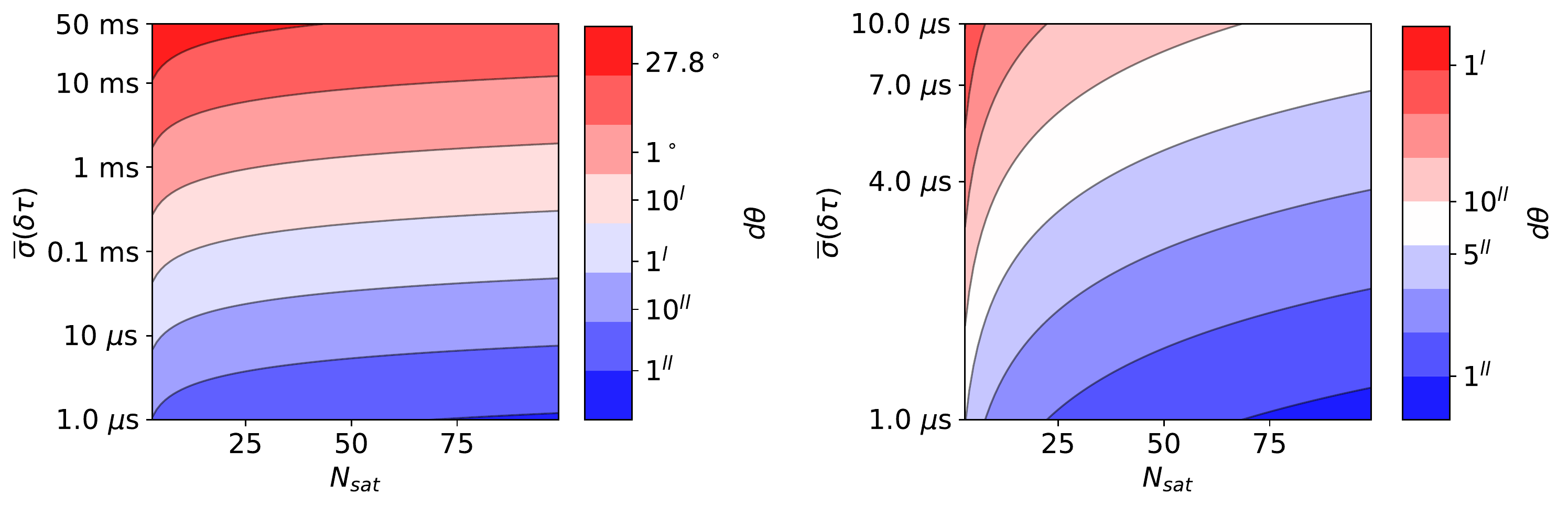}}
\caption{The left panel shows the relationship among mean uncertainty in the time delay of the satellites (log-scale) for the range between 1 $\mu$s and 50 ms, mean uncertainty in position accuracy (log-scale), and average number of satellites. The right panel shows the same relationship for mean uncertainty in the time delay among the satellites between 1 and 10 $\mu$s. The accuracies are calculated using Equation~\ref{eq:Gen}.}
\label{fig:CubeSatNumAccur}
\end{center}
\end{figure}

Using Equation~\ref{eq:Gen}, we calculated the position accuracies that can be achieved with a given number of satellites, which ranges from 3 to 100 ($N_{sat}$), and their average time delay accuracies, $\overline{\sigma}(\delta t)$, change between 1$\mu$s and 50 ms. The radius of Earth ($r_E$) is taken as 6378.1 km, while the orbital altitude for each satellite is assumed to be 580 km. Satellites at this altitude have orbital lifetimes typically in the order of $\sim$15-20 years \citep{Oltrogge2011AnEO}, before being incinerated in Earth's atmosphere due to its gravitational pull and net drag forces.

The results show that position accuracies larger than 1$^{\circ}$ can be achieved with less than 25 CubeSats (1$\sigma$ uncertainty from \citep{2013ApJS..207...39H}), with an average delay accuracy longer than 1 ms (left panel of Figure~\ref{fig:CubeSatNumAccur}). Reaching accuracies of 1${'}$ requires less than 20 CubeSats, with delay accuracies between around 0.1 ms and 10 $\mu$s, or more than 20 CubeSats with delay accuracies around 0.1 ms (left panel of Figure~\ref{fig:CubeSatNumAccur}). To zoom into accuracy range below 1${'}$ to meet the FOV requirement of the SONG telescope, we shortened the average time delay accuracies down to 1-10\,$\mu$s range and recalculated the relationship between localization accuracies and number of CubeSats (right panel of Figure~\ref{fig:CubeSatNumAccur}). The results show that around minimum 20 or less CubeSats with time delay accuracies range between 2-10\,$\mu$s provide localization accuracies between around 10${''}$ and 1${'}$, while better accuracies between around 5${''}$ and 1${''}$ require minimum 40 or more CubeSats with average time delay accuracies in the order of a few $\mu$s. However, it must be noted that reaching these accuracies is not achievable because of the intrinsic limitation of the SGRB light curves. Although sub-millisecond variations have been reported previously, it was generally accepted that millisecond variability is the the most common feature in the SGRB light curves \citep{10.1007/10853853_1}. 

The numbers calculated for the CubeSats in a mega-constellation to achieve high accuracy localization must also be multiplied by 2 as the half of the CubeSats in the constellation is expected to be in Earth's shadow for a uniformly distributed constellation. This results in $\sim$80 CubeSats for localization accuracies between a few 10 arcminutes and a few degrees. We must note that reaching these localization accuracies requires the uncertainty in the average time delay among the paired-CubeSats to be in the order of a few milliseconds.

%


\subsection{Communication Time Delay in Data Transmission}

Given the need for reaching a low latency in conveying event reports, we compute the communication time delay of our scenario. The delay for a transmission with acknowledgement during connection for a single path in $K$ number of hops in the network depends on the following:

\begin{itemize}
\item We consider a software configurable binning time of 5 ms for a data point of photon collection and total photon count reporting from readout electronics \citep{2005AandARau}. We consider that the light curve of a SGRB extends from 5 seconds before SGRB peak time to 5 seconds after it, considering the average SGRB burst duration of ~0.2 seconds \citep{2020ApJ...893...46V}. 
\item The data point byte representation of a measurement, as shown in Table~\ref{tab:segment}, since it will impact on the data segment size. To convey a measurement, we require: a timestamp with nanosecond temporal resolution (8 bytes), photon count value (2 bytes), latitude in degrees (4 bytes), longitude in degrees (4 bytes), orbit height in km (2 bytes), and spare bytes in case needed (5 bytes). Thus, a total of 25 bytes are required to represent a single measurement as a data point.
\item Based on the previous, we expect to send 2000 data points for the whole curve equivalent to 50 KB of data for a curve. Therefore, if two data points are sent per data segment, 1000 segments are required, and for 6 data points, at least 340 segments are needed. Smaller binning times are possible leading to larger number of segments. Therefore, we also consider larger number of segments in our results. Therefore, data segment sizes $s_d$ that includes 2 data points (50 bytes, referred as 50 B), and 6 data points (150 bytes, referred as 150 B) will be considered in our analysis.
\item The maximum number of satellites to transverse in this type of constellation, which coincides with the maximum number of hops, is $K_{max}=P+\floor{S/2}$ \citep{yang2012use}. In this case, the segments travel across all planes and then one half inside the final plane to reach a ground station.
\item The transmission time costs for data and acknowledgements, which are $s_d/R$ and $s_a/R$, respectively. $s_d$, $s_a$ and $R$ are the data segment size, acknowledgement segment size and the bit rate, respectively. 
\item Constant propagation time from forward and return paths, $2t_{p,i}$, where $t_{p, i}=d_{i}/c$ and $i=1,\ldots, K$, where $d_{i}$ are the propagation distances across hops that are intra-plane, inter-plane, and CubeSat to the ground station,
\item For each link, we consider working in regime known as propagation dominated, where the bit rate $R$ is larger than a threshold given by $R_{th, i}=(s_d - s_a)/2t_{p,i}\  \forall i$ \citep{hernandez2019delay}.
\end{itemize}

The delay for a transmission with acknowledgement during connection also depends on buffer queueing and internal device processing. However, we did not include these two processes as they are negligible compared to the delays stemming from propagation and transmission processes for small embedded devices \citep{paramanathan2014sharing,2016hernandezgoodput}.

\begin{table}[h]
\centering
\begin{threeparttable}
\caption{\label{tab:segment}Data point representation.}
\begin{tabular}{cc}
\hline
\hline
Variable & Byte size (B) \\
\hline
Timestamp (ns) & 8 \\
Photon count value & 2 \\
Latitude (degrees) & 4 \\
Longitude (degrees) & 4 \\
Orbit height (km) & 2 \\
Spare & 5 \\
\hline
\end{tabular}
\end{threeparttable}
\end{table}

Hence, we calculate the total mean communication delay, which includes both the cost rates for data transmission and the acknowledgements for a propagation dominated, separately, based on the following; 

\begin{eqnarray}
\label{eq:rlnc_delay1}
\mean{T_{d}} = \sum_{i = 1}^{K}\left(t_{p,i} + \frac{s_{c}}{R}\right) + \frac{Ks_{c}}{R}\mean{N_{nf}}\end{eqnarray}

\begin{eqnarray}
\mean{T_{a}} = \sum_{i = 1}^{K}\left(t_{p,i} + \frac{s_{a}}{R}\right)
\label{eq:rlnc_delay2}
\end{eqnarray}

\noindent where $\mean{T_{d}}$ and $\mean{T_{a}}$ represent delays from the data transmission and the acknowledgements, respectively. We also modify the transmitted segment size as $s_c = s_d + \alpha$ to account for the overhead of the coding coefficients. The delay for RLNC is then given by the sum of the data and acknowledgement delays defined in Equations~\ref{eq:rlnc_delay1} and \ref{eq:rlnc_delay2}. The term $\mean{N_{nf}}$ in Equations~\ref{eq:rlnc_delay1} and~\ref{eq:rlnc_delay2} represents the mean number of transmissions of RLNC without acknowledgements in a single path with $K$ hops, and is equal to $\mean{T_{nf}}/(1\,~s)$. $\mean{T_{nf}}$ is a unitary delay which is given in \citep{dikaliotis2014delay},

\begin{eqnarray}
\label{eq:unit_delay}
\mean{T_{nf}}(g, K, \epsilon_1,\ldots,\epsilon_K) &= \frac{\mean{N_{w}}}{1 - \epsilon_{w}} + \sum_{i=1,i \neq w}^{K} \frac{\epsilon_{w}}{\epsilon_{w} - \epsilon_i}
\end{eqnarray}

\noindent where $\mean{N_{w}}$ is the mean number of transmissions without losses in the worst hop, $\epsilon_i$ is the segment loss rate for each hop and $\epsilon_w$ is the worst segment loss rate in the network. The segment loss rates for each hop are given as $\epsilon_i = 1 - (1 - p_{B,i})^{s}$, where $i=1,\ldots, K$ and we consider independent and identically distributed bit errors in a received segment. It must be noted that we ensure that the system operates above an energy-per-bit to spectral noise density threshold for each hop to achieve an operational BER, $p_{B,i}$ in link $i$. The term $p_{B,i}$ is the bit error rate (BER), defined as the resulting rate of the number of bit errors divided by total bits sent in a transmission on average, and $s$ is the total transmitted segment size in bits for that given link.

\begin{figure}[htb!]
\begin{center}
{\includegraphics[width=2.75in]{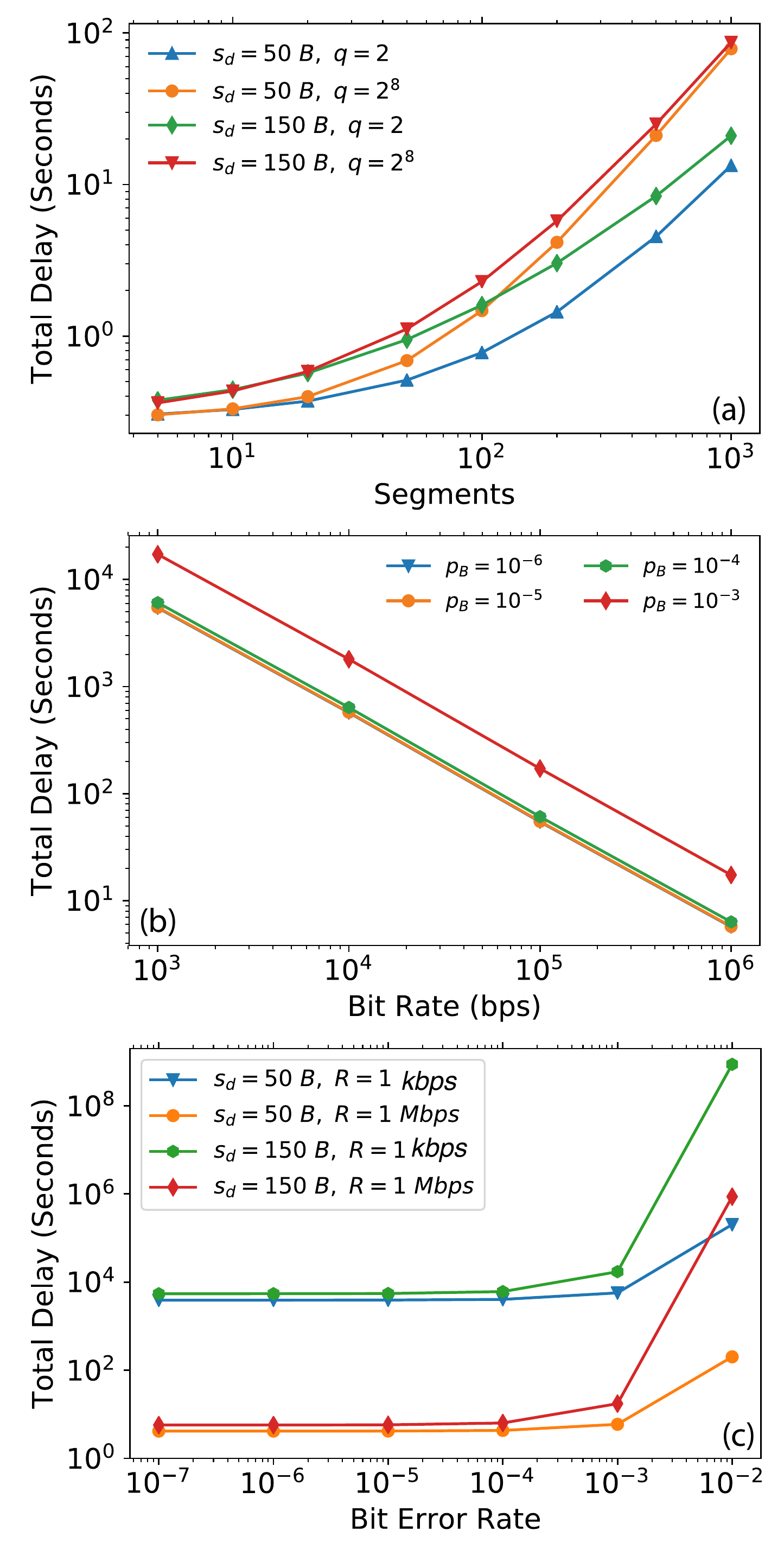}}
\caption{Relationship between the total delay and (a) number of segments for data segment sizes of 50 bytes and 150 bytes, and GF sizes of $q=2$ and $q=2^{8}$, (b) bit rate for BERs ranging from 10$^{-6}$ to 10$^{-3}$, $q = 2^8$; and (c) BER for data segment sizes of 50 bytes and 150 bytes, $q = 2^8$, and bit rates of 1 kbps and 1 Mbps. $s_a$ = 20 bytes in all cases.}
\label{fig:delay_segmentnum}
\end{center}
\end{figure}

We first study the relationship between the total delay and number of segments for data segment sizes of 50 bytes and 150 bytes, and GF sizes of $q=2$ and $q=2^{8}$ (Figure~\ref{fig:delay_segmentnum}a). For this calculation, we assume that we have 4 polar orbital planes, 10 CubeSats in each plane, minimum elevation angle of $\varepsilon_{min} = 0^{\circ}$, bit rate of $R = $ 1 Mbps, and BER $p_{B,i} = p_B = 10^{-5}$ for all links in the constellation. Having 40 CubeSats in 4 orbital planes is expected to provide localization accuracies between around 10${'}$ and 1${^\circ}$. 

The results show that the total communication delay increases as the number of segments increase for all data segments and GF sizes. Given our current assumptions, for bin sizes as 5 ms, a readout in the order of $10^3$ photons would lead a 1000 segments to be sent. This corresponds to the longest total delay of $\sim$90 s, which is calculated for GF of $q=2^{8}$ for the both data segment sizes (red and orange lines in Figure~\ref{fig:delay_segmentnum}a) while for larger bin sizes (as our reference value of 50 ms) the shortest total communication delay of $\sim$2 s would apply for GF of $q=2$ and data segment size of 50 bytes (blue line in Figure~\ref{fig:delay_segmentnum}a). In general, for a large number of segments, the smaller GF size results in shorter total delay, regardless of the data segment sizes for the assumed orbital configuration, because smaller field require less bits to represent the coding coefficients that are appended to a coded segment. For the case of GF of $q=2$, only one bit per coding coefficient is necessary to signal it, whereas for GF of $q=2^8$ one byte per coding coefficient is required that increase the number of bits to be transmitted and hence the total delay. On the other hand, for a number of segments below 100, the shortest time delay is calculated for the smallest data segment size (blue and orange lines in Figure~\ref{fig:delay_segmentnum}a), while the largest data segment size results in longer time delays (red and green lines in Figure~\ref{fig:delay_segmentnum}a). Furthermore, for a small number of segments per generation, GF of $q=2$ results in longer delays than GF of $q=2^8$. In this case, the major contributing factor for the delay of GF of $q=2$ is not the coding coefficients, but the frequent transmission of linearly dependent coded segment, particularly at the end of the generation.

Similar to the relationship between total delay and number of segments, we use 4 orbital planes accommodating 10 CubeSats in each plane to study the relationship between the bit rates and total delay. Here, we assume that the data segment size is $s_{d}$ = 150 bytes, acknowledgement segment size is $s_{a}$ = 20 bytes, 200 segments to transmit, corresponding to 1200 data points if more resolution is needed, and GF of $q=2^{8}$. We choose these parameters as a representative example of a bulk transmission of payload data with considerable overhead from the protocol stack to treat bits indistinctly. We changed the BERs from 10$^{-6}$ to 10$^{-3}$ to study the effects of different BERs on the relationship between the total delay and bit rates (Figure~\ref{fig:delay_segmentnum}b), since practical BERs for radio links with these applications are in this range\footnote{\url{https://gomspace.com/shop/subsystems/communication-(1)/default.aspx}}. The power law relationship between the bit rates and total delay shows that for higher (lower) bit rates show shorter (longer) delays for all BERs considered given that the bits are sent into the channel more frequently. The overall levels of the power law relationships is higher for BERs of 10$^{-3}$, while they are lower with smaller BERs (Figure~\ref{fig:delay_segmentnum}b). For BERs 10$^{-5}$ and 10$^{-6}$, the power law relationship between the bit rates and total delay overlaps. Therefore, target BERs smaller than $10^{-4}$ do not provide further reductions in the delay.

We also calculate the total communication delay for BERs ranging from 10$^{-7}$ to 10$^{-2}$ for the same constellation configuration as in previous calculations. However, this time we consider four scenarios where the data segment sizes are 50 bytes and 150 bytes, and the bit rates are 1 kbps and 1 Mbps (Figure~\ref{fig:delay_segmentnum}c). The results show that smaller bit rates (1 kbps) cause almost 1000 times longer total delays compared to the larger bit rates (1 Mbps), regardless of the data segment size. For the bit rate of 1 kbps, larger data segments result in $\sim$1000 s longer delays than smaller data segments do for the BERs ranging from 10$^{-7}$ to around 10$^{-4}$. For larger BERs of 10$^{-2}$, the total communication delay for 50 bytes is in the order of $\sim10^5$ s, where as for 150 bytes goes close to four orders of magnitude higher (blue and green lines in Figure~\ref{fig:delay_segmentnum}c), which would be unfeasible in practice. The larger bit rates show the same behaviour, where larger data segment size causing longer delays (red and orange lines in Figure~\ref{fig:delay_segmentnum}c). Interestingly, BERs of around 5$\times$10$^{-2}$, the high bit rates of 1 Mbps with data segment size of 150 bytes causes longer delay than that lower bit rate of 1 kbps with data segment size 50 bytes results in (red and blue lines in Figure~\ref{fig:delay_segmentnum}c). This is caused by the relationship of the segment loss rate with the BER and segment size. Therefore, at moderate segment loss rates and fixed BERs, the segment size plays an important role in controlling the total delay. Thus, we consider that data rates on the order of kbps should not be used for communications, where data rates in the order of Mbps provide relatively good results.

\begin{figure}[htb!]
\begin{center}
{\includegraphics[width=2.75in]{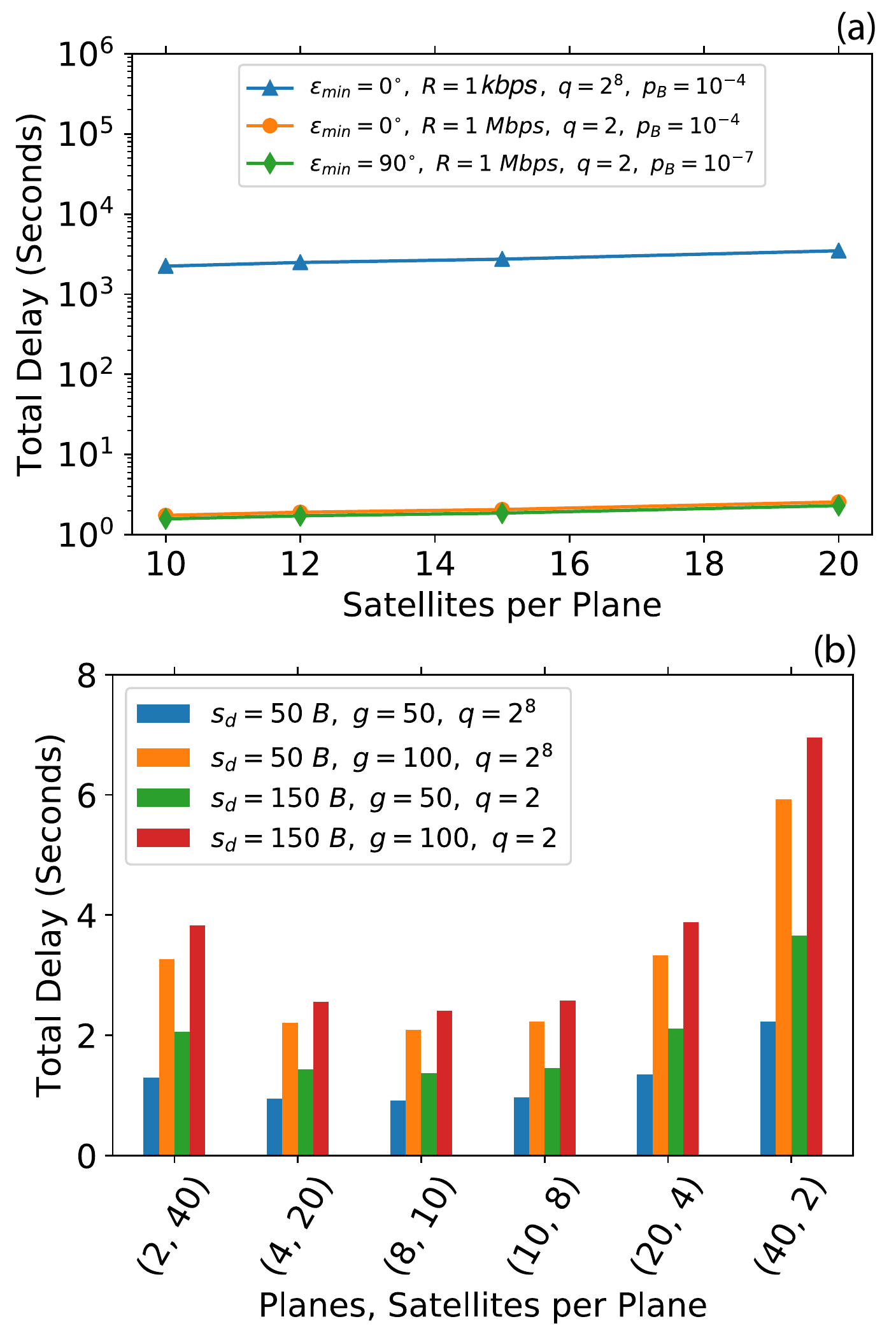}}
\caption{The top panel shows the relationship between the total delay and number of satellites for 4 planes where 100 segments are transmitted. The bottom panel shows the interplay between number of planes and CubeSats per plane for a total number of 80 CubeSats.}
\label{fig:delay_satnum}
\end{center}
\end{figure}

We compute the total time delay given by the sum Equations~\ref{eq:rlnc_delay1} and \ref{eq:rlnc_delay2}, extending from the detection of a SGRB to reception of this signal by a ground station, for a number of satellites in a mega-constellation of CubeSats, IMPACT. To achieve this, we chose best- and worst-case scenarios. For the best-case scenario, we assumed the elevation angle is 90$^{\circ}$, data segment size is $s_{d}$ = 50 bytes, the bit rate is 1 Mbps, GF of $q=2$, and BER of 10$^{-7}$. For the worst-case scenario, these parameters are taken as 0$^{\circ}$, $s_{d}$ = 150 bytes, the bit rate is 1 kbps, GF of $q=2^8$, and BER = 10$^{-2}$ (top panel of Figure~\ref{fig:delay_satnum}). We also include two more realistic scenarios where we fixed the data segment size to $s_d$ = 150 bytes, an elevation angle of 0$^\circ$, and BER of 10$^{-4}$ to achieve more feasible link budgets. The two scenarios include; (i) bit rate 1 kbps, GF of $q=2^{8}$, and BER of 10$^{-4}$, and (ii) bit rate of 1 Mbps, and GF of $q=2$. Our scenarios improve the delay bounds for a more realistic setup of the communication parameters since we consider a more practical operational BER. Our best-case scenario is close within a small relative margin to the best scenario achieving a total delay of $\sim$ 2 s, the only difference being the elevation angle and segment size.

Finally, we investigate on the ideal constellation design parameters for this use case, particularly number of planes and satellites per plane. To achieve this, we keep fixed the total number of satellites to $N_{sat}=80$ and vary the number of planes $P$ and satellites per plane $S$ for each of the given set of communication parameters. For all constellations, we fix the elevation angle to 0$^{\circ}$, the bit rate to 1 Mbps and the BER to 10$^{-4}$, as these are realistic link performance values and orbital configurations to achieve our maximum SGRB localization accuracies between around 10${'}$ and 1${^\circ}$. The orange and green bars in the bottom panel of Figure~\ref{fig:delay_satnum} are our ideal transmission schemes considering the delay of the longest possible established route between a detecting CubeSat in the constellation and a ground station, through a given routing protocol. In practice, we exclude the case of one plane since it leads to ambiguity in localization of the SGRBs. We observe that an ideal set of orbital parameters for an orbital altitude of 580 km exist for all transmission schemes which is 8 planes and 10 satellites per plane. We also notice that the ideal orbital configuration is independent of the total amount of data to be delivered and the communication parameters utilized. However, this minimum will be sensitive to the BER, and bit rate assumptions on the intra-and inter-satellite links and the downlink to a ground station since the associated delay costs would change as well.

\section{Discussion and Conclusions}
\label{sec:conc}

After the detection of the gravitational wave signals in 2015 \citep{2016PhRvL.116f1102A}, and the later detection of a short gamma-ray burst by the Fermi telescope in 2017 after GW from a NS-NS binary \citep{2017PhRvL.119p1101A}, these evidences for the association of SGRBs with GWs started the era of gravitational wave astronomy. Therefore, the prompt localization of these phenomena will allow ground and space based observatories to observe and analyze faster source afterglows that will provide insights into astrophysics, dense matter, gravitation, and cosmology \citep{2017PhRvL.119p1101A}.

In this study, we calculate the number of CubeSats in a mega-constellation to achieve localization accuracies between around 5${''}$ and 1${''}$. As complementary missions to the bigger space-based gamma-ray observatories, CubeSats can provide smaller, cheaper, and solutions with faster deployment times. The overarching idea is to have a fully coupled, automated localization and observation system, which relies on a CubeSat mega-constellation for prompt localization; and on the Stellar Observations Network Group (SONG) telescope FOV, which is 30$''$\,$\times$\,20$''$. As an automated network of observatories, the SONG telescope can perform photometric and spectroscopic measurements of the afterglows. 

In triangulation based localizations, uncertainties in the time delays between each pair of CubeSats are the major contributors for accuracies in localization calculations. This introduces constraints on the selection of gamma-ray detector materials and their read-out electronics. The most promising gamma-ray detectors, that can provide minimum uncertainties in time information, are LaBr$_{3}$(Ce) scintillators. These provide the shortest decay time of 16 ns (Table~\ref{tab:Sint}) as well as 40\% and 17\% higher light yields compared to NaI(Tl) and CsI(Tl) crystals, respectively. A good choice for read-out electronics to LaBr$_{3}$(Ce) can be Silicon photomultipliers (SiPMs) as they are not sensitive to magnetic fields, they have high multiplication gain ($\sim$10$^{6}$) leading to negligible electronic noise, low operation voltages, compact designs, and they provide very high time resolution in the order of a few nanoseconds \citep{2015JPhCS.620a2001J,2018ITNS...65..645C,Butt2015}. An effective area of 400 cm$^{2}$ for the energies ranging from 50 to 300 keV for this detector will also provide high counting rates, and hence smaller uncertainties in the detection time.

We calculated that for a first order approximation, the number of CubeSats in a mega-constellation is $\sim$80 for localization accuracies between a few 10 arcminutes and a few degrees, and requires uncertainties in average time delays in the order of a few milliseconds. These numbers are calculated for the orbital altitude of 580 km. We also calculated the number of CubeSats for different orbital altitudes, as increase in the baseline between each detector pairs will in turn increase the localization accuracy. However, for a first order approximation, small changes in the orbital altitudes of around a $\sim$100 km will not affect the average number of satellites.

Additionally, we calculated the total time delay between the reception of the SGRB signal by the CubeSats and reception of the information by a ground station. A mega-constellation of CubeSats in the LEO is a multi-hop network, where we studied various communication aspects as the bit rate, BER, segment size and erasure correcting code to reduce the delay. To achieve the maximum SGRB localization accuracies between around 5${''}$ and 1${''}$, we keep the total number of satellites of 80 constant and change the number of planes and satellites per plane for each of the given set of communication parameters, that are data segment sizes of $s_d$ = 50 bytes and $s_d$ = 150 bytes, acknowledgement size of $s_a$ = 20 bytes, the bit rate of 1 Mbps and the BER $p_B=10^{-4}$. The resulting BER is within the range of target BER values for typical communication link design in CubeSat missions\footnote{\url{http://www.amsatuk.me.uk/iaru/spreadsheet.htm}}. The results showed that, for the given set of parameters, 8 planes each accommodating 10 satellites provides the shortest delay time of 5 s.

The BATSE detectors on the {\it Compton Gamma-Ray Observatory} calculated coordinates with an accuracy of $\pm$10$^{\circ}$ with a time delay of 5.5 s \citep{Barthelmy1994}. The Burst Alert Telescope on the Neil Gehrels {\it Swift} Observatory provides localization accuracies between 1--3${'}$ and the position is distributed within around 20 s of the initial detection\footnote{\url{https://swift.nasa.gov/proposals/tech\textunderscore appd/swiftta\textunderscore v14.pdf}}. Additionally, the GBM on the {\it Fermi Gamma-Ray Space Telescope} downlinks the burst notification in 5 seconds\footnote{\url{https://fermi.gsfc.nasa.gov/science/parameters.html}}. The delay time between the receipt of a SGRB signal and downlinking the data to the ground station for high-accuracy localization for IMPACT is 5 s, excluding ground processing and telescope movement, which is comparable with the bigger and more costly missions.

In conclusion, an interconnected multi-hop array of CubeSats for transients, IMPACT, will enable us to detect, localize and study the SGRBs as counterparts to GWs in an energy range ranging from 50-300 keV. IMPACT is planned to consist maximum 80 CubeSats, which are synchronized with an on-board GPS. These CubeSats are planned to be distributed in 8 orbital planes, each of which accommodating 10 CubeSats that carry LaBr$_{3}$(Ce) scintillation crystals coupled with SiMPs as gamma-ray detectors. Our calculations suggest that IMPACT will be able to provide localization accuracies between 10 arcminutes and 1$^{\circ}$, which requires uncertainties in average time delays in the order of a few milliseconds with 80 CubeSats. This orbital architecture will also provide an all sky coverage (the field of view is 4$\pi$ steradians). Additionally, the time it takes for IMPACT to detect a SGRB and downlink the required information for localization to a ground station is expected to be around 5 s.

\section{Additional Requirements}

For additional requirements for specific article types and further information please refer to \href{http://www.frontiersin.org/about/AuthorGuidelines#AdditionalRequirements}{Author Guidelines}.

\section*{Conflict of Interest Statement}

The authors declare that the research was conducted in the absence of any commercial or financial relationships that could be construed as a potential conflict of interest.

\section*{Author Contributions}
All authors collaborated in the motivation, conceptual design, methodology, results analysis and conclusions. FI reviewed the State-of-the-Art, SGRB scientific missions, CubeSats missions, detector payload evaluation, trilateration, number of CubeSats calculation, corresponding plotting and results. NJHM provided the communication payload State-of-the-Art, constellation, communication and erasure code delay evaluation, corresponding plotting and results. FI and NJHM discussed the results and conclusions with RHJ and CK. FI and NJHM wrote the paper, where RHJ and CK contributed on the manuscript providing significant observations to the study, results and conclusions.


\section*{Funding}
MegaMan project (J. nr. 7049-00003B).

\section*{Acknowledgments}
This work was funded by Innovation Fund Denmark as part of the MegaMan project (J. nr. 7049-00003B).



\bibliographystyle{frontiersinSCNS_ENG_HUMS} 
\bibliography{bibliography_numsat}

\end{document}